\documentclass[11pt,fleqn]{article}
\usepackage[]{fontenc}
\usepackage{amssymb}
\usepackage[latin1]{inputenc}
\usepackage{times}
\usepackage{mathptm}
\newcommand{\dagg}{{\scriptscriptstyle\dagger}}

\begin{document}

\title{Stability of Ferromagnetism in Hubbard models with degenerate
single-particle
ground states}

\author{Andreas Mielke\thanks{
E--mail: mielke@tphys.uni-heidelberg.de
}\\
 Institut f\"{u}r Theoretische Physik, Ruprecht--Karls--Universit\"{a}t, \\
 Philosophenweg 19, D-69120~Heidelberg, F.R.~Germany}

\maketitle
\begin{center}
Accepted for publication in J. Phys. A
\end{center}
\begin{abstract}
A Hubbard model with a \( N_{d} \)-fold degenerate single-particle ground state
has ferromagnetic ground states if the number of electrons is less or equal
to \( N_{d} \). It is shown rigorously that the local stability of
ferromagnetism
in such a model implies global stability: The model has only ferromagnetic
ground
states, if there are no single spin-flip ground states. If the number of
electrons
is equal to \( N_{d} \), it is well known that the ferromagnetic ground state
is unique if and only if the single-particle density matrix is irreducible.
We present a simplified proof for this result.\\
\\
 PACS-numbers: 75.10.Lp, 71.10.Fd
\end{abstract}

\section{Introduction }

\label{s1} The problem of ferromagnetism in itinerant electron systems has
a long history. It is clear that ferromagnetism (as any other ordering in
itinerant
electron systems) occurs due to the interaction of the electrons, or, to be
more precise, due to a subtle interplay between the kinetic motion of the
electrons
and the interaction. 1963, Hubbard \cite{Hubbard}, Kanamori \cite{Kanamori},
and Gutzwiller \cite{Gutzwiller} formulated and studied a simple tight-binding
model of electrons with an on-site Coulomb repulsion of strength \( U \). This
model is usually called the Hubbard model. Although the assumption, that a
realistic
system can be described by a purely local repulsion of the electrons is
artifical,
the Hubbard model became a paradigm for the study of correlated electron
systems.
The reason is that already a pure on-site interaction can produce many ordering
effects that have been observed in electronic systems. The mechanisms that are
responsible for some long range order in the ground state of the Hubbard model
are probably the same in more complicated (and more realistic) models. From
a theoretical point of view the Hubbard model is very interesting, because it
offers the possibility to derive ordering phenomena in a simple model that does
not contain special interactions favouring this order.

In this paper we present a result on ferromagnetism in the Hubbard model. This
is an old problem, which has been studied extensively using various
approximative
methods. The most simple approach is the Hartree-Fock approximation. It yields
the Stoner criterion \( U\rho _{F}>1 \) for the occurence of ferromagnetism
in the Hubbard model. \( \rho _{F} \) is the density of states at the fermi
energy. It is well known that this criterion overestimates the occurence of
ferromagnetism. There are situations where \( \rho _{F}U \) is infinite and
the ground state of the Hubbard model is not ferromagnetic. Ferromagnetism is
not a universal property of the Hubbard model. As far as we know it occurs on
special lattices and in special regions of the parameter space.

In the discussion of ferromagnetism in correlated electron systems, more
realistic
models with e.g. an additional ferromagnetic interaction between the elctrons
\cite{Wahle} or a Hund's coupling between several bands in a multi-band system
\cite{Kubo} have been discussed as well. It is clear that in a realistic
description
of itinerant ferromagnetism such additional interactions are present and may
favour the occurence of ferromagnetism. But it is a challenging problem to
derive
conditions for the occurence of ferromagnetism in a Hubbard model, which does
not explicitely contain such interactions. The hope is that results for this
model yield a main contribution to the understanding of ferromagnetism in more
realistic models.

The first rigorous results on ferromagnetism in the Hubbard model is the so
called Theorem of Nagaoka\cite{Nagaoka}. On a large class of lattices, the
Hubbard model has a ferromagnetic ground state if the Coulomb repulsion is
infinite
and if there is one electron less than lattice sites. A very general proof of
this theorem has been given by Tasaki \cite{Tasaki89}.

A second class of systems, for which the existence of ferromagnetic ground
states
has been shown rigorously, are the so called flat band models. In 1989, Lieb
\cite{Lieb89} proved an important theorem on the Hubbard model: At half filling
and on a bibartite lattice (one electron per lattice site) the ground state
of the Hubbard model is unique up to the usual spin degeneracy. The spin of
the ground state is given by \( S=\frac{1}{2}||A|-|B|| \), where \( |A| \)
and \( |B| \) are the numbers of lattice sites of the two sublattices of the
bipartite lattice. When this quantity is extensive, the system is
ferrimagnetic.
In that case, the model has strongly degenerate single particle eigenstates
at the Fermi level, \( \rho _{F} \) is infinite. For a Hubbard model on a
translational
invariant lattice such a model has several bands, one of these bands is flat.
Later, it has been shown that a multiband Hubbard model for which the lowest
band is flat shows ferromagnetism\cite{Mielke91a}-\cite{MielkeTasaki}. These
lattices are not bipartite, generally they contain triangles or
next-nearest-neighbor
hoppings. A typical example is the Hubbard model on the kagomé lattice
\cite{Mielke92}.

There are several extensions of the flat band ferromagnetism. The most
important
result has been derived by Tasaki \cite{Tasaki96}. He discussed the question,
whether the flat band ferromagnet is stable with respect to small
perturbations.
He showed under very general asumptions that for a class of multi-band Hubbard
models with a nearly flat lowest band the ferromagnetic state is stable with
respect to single spin flips if the Coulomb repulsion \( U \) is sufficiently
large and if the nearly flat band is half filled. This local stability of the
ferromagnetic state suggests its global stability. The class of models, for
which Tasaki was able to proof this important result consists of models, for
which the nearly flat, lowest band is separated from the rest of the spectrum
by a sufficiently large gap. Therefore one would expect that these models
describe
an insulating ferromagnet. This a general problem for the flat band
ferromagnetism
as well. The flat band models show ferromagnetism if the flat band is half
filled
or less than half filled. Even if the flat band is less than half filled, or
if the model has no gap between the flat band the other bands (this is the case
for the kagomé lattice), the system may be an insulator. The reason is that
for an entire flat band, the systme can be described by localized states as
well. Furthermore, the existence of a basis of localized states was an
essential
part of the proofs.

Another extension of the flat band ferromagnetism are models with a partially
flat band. In \cite{Mielke93} a general necessary and sufficient condition
for the uniqueness of ferromagnetic ground states has been derived for a model
with a degenerate single particle ground state. This result holds only if the
number of electrons is equal to the number of degenerate single particle ground
states. But it does not require a gap in the spectrum or an entirely flat band.
A partially flat band is sufficient. A Hubbard model with a single band far
away from half filling is expected to be a conductor. This remains true if the
band is partially flat. Therefore these models may describe a metallic
ferromagnet.
A generalization of this result to a situation where the number of electrons
is less than the number of single particle states has recently been published
\cite{Mielke99}. The main result of that letter is that in a single band
Hubbard
model with a degenerate single partical ground state local stability of
ferromagnetism
implies global stability, if the number of electrons is less or equal to the
number of degenerate single particle groud states. Stability is meant here in
the sence of absolute stability: The ferromagnetic ground state is the only
ground state of the system.

That local stability of ferromagnetism implies global stability has often been
assumed but is by no means guarantied. It would be certainly useful to know,
in which situatons this is the case. The aim of the present paper is to
generalize
the result in to a general Hubbard model with degenerate single particle ground
states. It is not necessary to have a single band model. It is even not
necessary
to have translational invariance, although this would be a natural asumption.

Let us mention that Hubbard models with a partially flat band are not only an
academic toy model. Very recently Arita et al \cite{Arita} used such a model
to explain the negative magnetoresistance of certain organic conductors. They
mention that standard band-structure calculations for these materials yield
a partially flat band.

The paper is organized as follows. The next section contains the main
definitions
and results. The proof of the result combines the ingredients of the proofs
in \cite{Mielke93} and. The main part of the proof is the choice of a suitable
basis. This choice is discussed in Section 3. Section 4 contains a new proof
of the result in \cite{Mielke93}. The proof in \cite{Mielke93} used an
induction
in the number of degenerate single particle states and was not intuitive at
all. On the other hand, the basic idea why the condition in \cite{Mielke93}
should be true is simple. The new proof is based on this basic idea and is much
easier. Furthermore it can be generalized to situations where the number of
electrons is less than the number of degenerate single particle states. This
generalization is presented in Section 5.

\section{Main result}

We consider a Hubbard model on a finite lattice with \( N_{s} \) sites. The
Hamiltonian is
\begin{equation}
H=H_{\textrm{hop}}+H_{\textrm{int}},
\end{equation}
 where
\begin{equation}
H_{\textrm{hop}}=\sum _{x,y,\sigma }t_{xy}c_{x\sigma }^{\dagger }c_{y\sigma },
\end{equation}
 and
\begin{equation}
H_{\textrm{int}}=\sum _{x}U_{x}n_{x\uparrow }n_{x\downarrow }.
\end{equation}
 \( x \) and \( y \) are lattice sites. As usual \( c_{x\sigma }^{\dagger } \)
and \( c_{x\sigma } \) are the creation and the annihilation operators of an
electron on the site \( x \) with a spin \( \sigma =\uparrow ,\downarrow  \).
They satisfy the anticommutation relations \( \left[ c_{x\sigma },c_{y\tau
}^{\dagger }\right] _{+}=\delta _{xy}\delta _{\sigma \tau }, \)
and \( \left[ c_{x\sigma },c_{y\tau }\right] _{+}=\left[ c_{x\sigma }^{\dagger
},c_{y\tau }^{\dagger }\right] _{+}=0. \)
The number operator is defined as \( n_{x\sigma }=c_{x\sigma }^{\dagger
}c_{x\sigma }. \)
The hopping matrix \( T=(t_{xy}) \) is real symmetric, and the on-site Coulomb
repulsion \( U_{x} \) is positive. We do not need to assume any kind of
translational
symmetry, therefore the lattice is simply a collection of sites. We allow that
the local Coulomb repulsion depends on \( x \). The total number of electrons
is \( N_{e}=\sum _{x\in \Lambda }(n_{x\uparrow }+n_{x\downarrow }) \).

The Hubbard model has a \( SU(2) \) spin symmetry, it commutes with the spin
operators
\begin{equation}
\overrightarrow{S}=\sum _{x}\sum _{\sigma ,\tau =\uparrow ,\downarrow
}c^{\dagger }_{x\sigma }(\overrightarrow{p})_{\sigma \tau }c_{x\tau }/2,
\end{equation}
 where \( \overrightarrow{p} \) is the vector of Pauli matrices,
\begin{equation}
\label{Pauli}
p_{1}=\left( \begin{array}{cc}
0 & 1\\
1 & 0
\end{array}\right) ,\quad p_{2}=\left( \begin{array}{cc}
0 & -i\\
i & 0
\end{array}\right) ,\quad p_{3}=\left( \begin{array}{cc}
1 & 0\\
0 & -1
\end{array}\right) .
\end{equation}
 We denote by \( S(S+1) \) the eigenvalue of \( \overrightarrow{S}^{2} \).

The eigenstates of the hopping matrix \( T \) are \( \varphi ^{i} \), the
corresponding eigenvalues are \( \epsilon _{i} \), \( i=1,\ldots ,N_{s} \).
Let \( \epsilon _{i}\leq \epsilon _{j} \) for \( i<j \). In the following
we will discuss a Hubbard model with \( N_{d} \) degenerate single particle
ground states. The energy scale is chosen such that \( \epsilon _{i}=0 \) for
\( i\leq N_{d} \).

The main result of the present paper can now be formulated.

\begin{quote}
\emph{Theorem} -- In a Hubbard model with \( N_{d} \) degenerate single
particle
ground states and \( N_{e}\leq N_{d} \) electrons, local stability of
ferromagnetism
implies global stability: The model has only ferromagnetic ground states with
a spin \( S=\frac{N_{e}}{2} \), if and only if there are no single spin-flip
ground states (ground states with a spin \( S=\frac{N_{e}}{2}-1 \)).
\end{quote}
This theorem is true for any positive Coulomb repulsion \( U_{x} \). The
existence
of feromagnetic ground states is indeed trivial. Any multi particle state that
contains only electron with spin up in single particle states \( \varphi ^{i}
\),
\( i\leq N_{d} \) is a ground state of the Hamiltonian. It is even a ground
state of the kinetic part and of the interaction part of the Hamiltonian
separately.
The problem is to show that there are no further ground states. The above
theorem
yields a necessary and sufficient condition for the existence of
non-ferromagnetic
ground states.

As already mentioned in \cite{Mielke93}, one can use degenerate perturbation
theory to generalize the result to a situation where the flat part of the band
lies not at the bottom of the single-particle spectrum. A concrete model, where
such a situation occurs, has been investigated by Arita et al \cite{Arita98}.
They investigated a special type of one-dimensional Hubbard model used for the
description of atomic quantum wires. These models have a flat band that is not
situated at the bottom of the spectrum.

\section{Choice of the basis}

The main part of the proof of the theorem is the choice of an appropriate
basis.
It turns out that the choice of the single particle basis used in
\cite{Mielke93}
is useful. In this section we give a detailed construction of such a basis.
Starting point is the representation

\begin{equation}
T=\left( \begin{array}{cc}
C^{\dagg }T_{0}C & C^{\dagg }T_{0}\\
T_{0}C & T_{0}
\end{array}\right)
\end{equation}
of the hopping matrix. Here \( T_{0} \) is a positive \( (N_{s}-N_{d})\times
(N_{s}-N_{d}) \)-matrix,
\( {\textrm{rank}}T_{0}=N_{s}-N_{d} \). \( C \) is a \( (N_{s}-N_{d})\times
N_{d} \)-matrix.
This representation of \( T \) can be obtained as follows: Since \(
{\textrm{rank}}T=N_{s}-N_{d} \),
one can find \( N_{s}-N_{d} \) rows (or colums) of \( T \) wich are linear
independent. We label the corresponding sites by \( x=N_{d}+1,\ldots ,N_{s} \).
\( T_{0} \) is the submatrix \( (t_{xy})_{x,y\in \{N_{d}+1,\ldots ,N_{s}\}} \).
Since \( T \) is non-negative, \( T_{0} \) is positive. The matrix \( C \)
is given by \( (T_{0})^{-1}T_{01} \) where \( T_{01}=(t_{xy})_{x\in
\{N_{d}+1,\ldots ,N_{s}\},y\in \{1,\ldots ,N_{d}\}} \).
The other matrix elements of \( T \) are fixed, since \( T \) is symmetric
and since the other rows of \( T \) are linear dependent.

By construction, the single particle ground states obey \( T\psi =0 \). This
holds if and only if
\begin{equation}
\psi =\left( \begin{array}{c}
\bar{\psi }\\
-C\bar{\psi }
\end{array}\right)
\end{equation}
 A basis of single particle ground states can be obtained by choosing an
arbitrary
set of \( N_{d} \) linear independent vectors \( \bar{\psi } \). We choose
the basis \( {\mathcal{B}}=\{\psi _{i}:\bar{\psi }_{i}(x)=\delta _{x,i}\} \).
Since \( t_{xy} \) are real, \( \psi _{i}(x) \) are real. This basis is not
orthonormal. The matrix \( B=(b_{ij})_{i,j=1,\ldots ,N_{d}} \) with \(
b_{ij}=\sum _{x}\psi _{i}(x)\psi _{j}(x) \)
is positive and the dual basis is formed by \( \psi _{i}^{d}(x)=\sum
_{j}(B^{-1})_{ij}\psi _{j}(x). \)
One has \( \sum _{x}\psi _{i}^{d}(x)\psi _{j}(x)=\delta _{i,j} \). We introduce
creation operators for electrons in the state \( \psi _{i}(x) \)
\begin{equation}
a_{i\sigma }^{\dagg }=\sum _{x}\psi _{i}(x)c_{x}^{\dagg }
\end{equation}
 and the corresponding dual operators
\begin{equation}
a_{i\sigma }=\sum _{x}\psi _{i}^{d}(x)c_{x}^{\dagg }
\end{equation}
They obey the commutation relations \( [a_{i\sigma },a^{\dagg }_{j\tau
}]=\delta _{i,j}\delta _{\sigma ,\tau } \).
These creation and annihilation operators can now be used to construct
multi-particle
states. The unique ferromagnetic ground state with \( N_{e}=N_{d} \) electrons
and \( S=S_{3}=N_{d}/2 \) is
\begin{equation}
\psi _{0F}=\prod _{i}a^{\dagg }_{i\uparrow }|0\rangle
\end{equation}
 A general ground state of the kinetic part of the Hamiltonian is given by

\begin{equation}
\label{ansatz}
\psi ^{n,m}(\alpha )=S_{-}^{n,m}(\alpha )\psi _{0F}
\end{equation}
 where
\begin{equation}
S_{-}^{n,m}(\alpha )=\sum _{j_{1}\ldots j_{m};i_{1}\ldots i_{n}}\alpha
_{j_{1}\ldots j_{m};i_{1}\ldots i_{n}}\prod _{k}a^{\dagg }_{j_{k}\downarrow
}\prod _{k}a_{i_{k}\uparrow }
\end{equation}
This state has \( N_{e}=N_{d}-n+m \) electrons. In the following I assume that
\( N_{e}\le N_{d} \), i.e. \( m\le n \). \( \psi ^{n,m}(\alpha ) \) is a
state with \( S_{3}=(N_{d}-n-m)/2=N_{e}/2-m \). It obeys \( S_{+}\psi
^{n,m}(\alpha )=0 \)
if and only if
\begin{equation}
\sum _{k}\alpha _{k,j_{1}\ldots j_{m-1};k,i_{1}\ldots i_{n-1}}=0.
\end{equation}
 In that case it is a state with a spin \( S=S_{3} \). We want to derive a
condition for \( \psi ^{n,m}(\alpha ) \) to be a ground state of the
Hamitltonian.
A necessary and sufficient condition is
\begin{equation}
c_{x\uparrow }c_{x\downarrow }\psi ^{n,m}(\alpha )=0
\end{equation}
 for all \( x \). For \( x\le N_{d} \) one obtains \( a_{i\uparrow
}a_{i\downarrow }\psi ^{n,m}(\alpha )=0 \).
Therefore \( \alpha _{j_{1}\ldots j_{m};i_{1}\ldots i_{n}}=0 \) if \(
\{j_{1}\ldots j_{m}\} \)
is not a subset of \( \{i_{1}\ldots i_{n}\} \). It turns out that this fact
is important since it simplifies the proof substantially.

\section{The case \protect\( N_{e}=N_{d}\protect \)}

This case has already been discussed in \cite{Mielke93}. In the following we
obtain a simplified proof of the result in \cite{Mielke93}. Let us first
discuss
the stability with respect to single spin-flips. A ferromagnetic ground state
is called stable with respect to a single spin flip, if there is no single spin
flip state with the same energy. I derive a necessary and sufficient condition
for \( \psi _{0F} \) to be stable with respect to a single spin flip. A general
state with a single spin flip can be written in the form
\begin{equation}
\psi =\sum _{j,k}\alpha _{j;k}c^{\dagg }_{j\downarrow }c_{k\uparrow }\psi _{0F}
\end{equation}
 If and only if \( \alpha _{j;k}\propto \delta _{j,k} \), \( \psi  \) is the
unique ferromagnetic ground state with \( S=N_{d}/2 \), \( S_{z}=N_{d}/2-1 \).
If and only if \( \sum _{j}\alpha _{j;j}=0 \), \( \psi  \) is a state with
\( S=S_{z}=N_{d}/2-1 \). Therefore I assume \( \sum _{j}\alpha _{j;j}=0 \).
\( \psi  \) is a ground state if and only if \( c_{x\uparrow }c_{x\downarrow
}\psi =0 \).
\begin{equation}
c_{x\uparrow }c_{x\downarrow }\psi =\sum _{j,k,l}\alpha _{j;k}\psi _{j}(x)\psi
_{l}(x)c_{l\uparrow }c_{k\uparrow }\psi _{0F}
\end{equation}
 The right hand side vanishes if and only if
\begin{equation}
\sum _{j}\psi _{j}(x)(\alpha _{j;k}\psi _{l}(x)-\alpha _{j;l}\psi
_{k}(x))=0\quad \forall k,\, l
\end{equation}
 I introduce
\begin{equation}
\tilde{\psi }_{k}(x)=\sum _{j}\alpha _{j;k}\psi _{j}(x)
\end{equation}
 The condition for \( \alpha _{j,k} \) yields
\begin{equation}
\label{cond1}
\tilde{\psi }_{k}(x)\psi _{l}(x)-\tilde{\psi }_{l}(x)\psi _{k}(x)=0\quad
\forall k,\, l,\, x
\end{equation}
 A trivial solution is \( \tilde{\psi }_{k}(x)=\psi _{k}(x) \). It corresponds
to \( \alpha _{j,k}\propto \delta _{j,k} \) and has been excluded above.
Multiplying
the condition for \( \tilde{\psi }_{k}(x) \) with \( \psi ^{d}_{l}(y)\psi
^{d}_{k}(z) \)
and summing over \( k \) and \( l \) yields
\begin{equation}
\tilde{\rho }_{y,x}\rho _{x,z}-\rho _{y,x}\tilde{\rho }_{x,z}=0
\end{equation}
 where \( \rho _{y,x}=\sum _{j}\psi _{j}(x)\psi ^{d}_{j}(y) \), \( \tilde{\rho
}_{y,x}=\sum _{j}\tilde{\psi }_{j}(x)\psi ^{d}_{j}(y) \).
If the matrix \( \rho _{x,y} \) is irreducible, the only solution is \(
\tilde{\rho }_{y,x}=\rho _{y,x} \).
It corresponds to \( \alpha _{j,k}\propto \delta _{j,k} \) and has been
excluded
above. If the matrix \( \rho _{x,y} \) is reducible, the equation for \(
\tilde{\rho }_{y,x} \)
has another non-trivial solution. From the non-trivial solution for \(
\tilde{\rho }_{y,x} \)
one obtains a solution for \( \alpha _{j,k} \) from which one can construct
easily a solution with \( \sum _{j}\alpha _{j,j}=0 \). Thus \( \psi _{0F} \)
is stable with respect to a single spin flip if and only if \( \rho _{x,y} \)
is irreducible. This is the condition derived previously in \cite{Mielke93}.
To derive this condition, it was not necessary to use the special single
particle
basis introduced in Section 3. The use of this basis is useful for the
investigation
of multi spin-flip states.

Let us now consider a multi spin-flip state \( \psi ^{n,m}(\alpha ) \). It
is a ground state if and only if

\begin{equation}
\label{cond_nn}
\sum _{P}(-1)^{P}\sum _{j_{1}}\psi _{j_{1}}(x)\psi _{k_{P(n+1)}}\alpha
_{j_{1}\ldots j_{n};k_{P(1)}\ldots k_{P(n)}}=0
\end{equation}
 Since \( \alpha _{j_{1}\ldots j_{n};i_{1}\ldots i_{n}} \) is antisymmetric
in the last \( n \) indices, it is sufficient to sum over all cyclic
permutations.
\begin{equation}
\label{cond_nnc}
\sum _{r=1}^{n+1}(-1)^{nr}\sum _{j_{1}}\psi _{j_{1}}(x)\psi _{k_{r}}(x)\alpha
_{j_{1}\ldots j_{n};k_{r+1}\ldots k_{n+1},k_{1}\ldots k_{r-1}}=0
\end{equation}
Since \( \alpha _{j_{1}\ldots j_{n};i_{1}\ldots i_{n}}\ne0  \) only if \(
\{j_{1}\ldots j_{n}\}=\{i_{1}\ldots i_{n}\} \),
we obtain for \( n=1 \) (the single spin-flip case)
\begin{equation}
\psi _{k}(x)\psi _{k'}(x)(\alpha _{k;k}-\alpha _{k';k'})=0
\end{equation}
 With \( \tilde{\psi }_{k}(x)=\alpha _{k,k}\psi _{k}(x) \) this yields the
original condition (\ref{cond1}). This means, that with this choice of the
basis, the functions \( \tilde{\psi }_{k}(x) \) are either equal to \( \psi
_{k}(x) \)
or vanish. A solution exists, if the set \( \{\psi _{k}(x),k=1\ldots N_{d}\} \)
decays in two subsets such that \( \psi _{k}(x)\psi _{k'}(x)=0 \) if the two
factors are out of different subsets. This is equivalent to the above condition
on the single particle density matrix \( \rho _{x,y} \). For \( n>1 \) we
now use the fact that the set \( \{j_{1},\ldots ,j_{n}\} \) is a subset of
\( \{k_{1},\ldots ,k_{n+1}\} \). I choose \( j_{2}=k_{1} \), \( j_{3}=k_{2} \),
etc in (\ref{cond_nnc}). With this choice only the terms with \( r\ge n \)
in the sum over \( r \) do not vanish. For \( r=n \) the only non-vanishing
contribution in the sum over \( j_{1} \) is \( j_{1}=k_{n+1} \). For \( r=n+1
\)
one has \( j_{1}=k_{n} \). This finally yields
\begin{equation}
\psi _{k_{n}}(x)\psi _{k_{n+1}}(x)(\alpha _{k_{1}\ldots k_{n};k_{1}\ldots
k_{n}}-\alpha _{k_{n+1}k_{1}\ldots k_{n-1};k_{n+1}k_{1}\ldots k_{n-1}})=0
\end{equation}
 For some fixed \( k_{1},\ldots \, k_{n-1} \) I let \( \alpha _{k;k}=\alpha
_{kk_{1}\ldots k_{n-1};kk_{1}\ldots k_{n-1}} \).
The indices \( k_{1}\ldots \, k_{n-1} \) are chosen such that \( \alpha _{k;k}
\)
does not vanish identically, which is possible since \( \alpha _{j_{1}\ldots
j_{n};i_{1}\ldots i_{n}} \)
does not vanish identically. This shows that the existence of a multi spin-flip
ground state implies the existence of a single spin-flip ground states.
Therefore
the ferromagnetic ground state of the Hubbard model with \( N_{e}=N_{d} \)
electrons is the unique ground state (up to the spin degeneracy due to the \(
SU(2) \)
symmetry) if and only if \( \rho _{xy} \) is irreducible.

\section{The case \protect\( N_{e}<N_{d}\protect \)}

It is now very easy to generalize the second part of the above derivation to
the case \( N_{e}<N_{d} \). We will show that the existence of a multi
spin-filp
ground state implies the existence of a single spin-flip ground state. Let \(
\psi ^{1,n}(\alpha ) \),
\( n>1 \) be a single spin-flip ground state for \( N_{e}=N_{d}-n+1 \)
electrons.
The condition, that this is a ground state, yields
\begin{equation}
\label{cond_nm1}
\sum _{r=1}^{n+1}(-1)^{nr}\sum _{j_{1}\in \{k_{1},\ldots k_{n+1}\}\setminus
\{k_{r}\}}\psi _{j_{1}}(x)\psi _{k_{r}}(x)\alpha _{j_{1};k_{r+1}\ldots
k_{n+1}k_{1}\ldots k_{r-1}}=0
\end{equation}
 The sum over \( j_{1} \) is restricted to the set \( \{k_{1},\ldots
k_{n+1}\}\setminus \{k_{r}\} \)
since otherwise \( \alpha _{j_{1};k_{r+1}\ldots k_{n+1}k_{1}\ldots k_{r-1}} \)
vanishes. The similar condition for a multi spin-flip ground state \( \psi
^{m,n+m-1}(\alpha ) \)
is
\begin{equation}
\label{cond_nm}
\sum _{r=1}^{n+m}(-1)^{(n+m-1)r}\sum _{j_{1}\in \{k_{1},\ldots
k_{n+m}\}\setminus \{k_{r},j_{2}\ldots j_{m}\}}\psi _{j_{1}}(x)\psi
_{k_{r}}(x)\alpha _{j_{1}\ldots j_{m};k_{r+1}\ldots k_{n+m}k_{1}\ldots
k_{r-1}}=0
\end{equation}
I let \( j_{r}=k_{n+r} \), \( r\ge 2 \). Then the sum over \( r \) runs from
\( 1 \) to \( n+1 \) and the sum over \( j_{1} \) runs over all elements
of \( \{k_{1}\ldots k_{n+1}\}\setminus \{k_{r}\} \). all other terms vanish
identically. One obtains
\begin{equation}
\sum _{r=1}^{n+1}(-1)^{nr}\sum _{j_{1}\in \{k_{1},\ldots k_{n+1}\}\setminus
\{k_{r}\}}\psi _{j_{1}}(x)\psi _{k_{r}}(x)\alpha _{j_{1}k_{n+2}\ldots
k_{n+m};k_{r+1}\ldots k_{n+1}k_{1}\ldots k_{r-1}k_{n+2}\ldots k_{n+m}}=0
\end{equation}
 Therefore we can choose \( \tilde{\alpha }_{j_{1};k_{r+1}\ldots
k_{n+1}k_{1}\ldots k_{r-1}}=\alpha _{j_{1}k_{n+2}\ldots k_{n+m};k_{r+1}\ldots
k_{n+1}k_{1}\ldots k_{r-1}k_{n+2}\ldots k_{n+m}} \)
for some fixed \( k_{n+2},\ldots \, k_{n+m} \), such that \( \tilde{\alpha
}_{j_{1};k_{r+1}\ldots k_{n+1}k_{1}\ldots k_{r-1}} \)does
not vanish identically. This is possible since \( \alpha _{j_{1}\ldots
j_{m};k_{1}\ldots k_{n}} \)
does not vanish identically. The corresponding single spin-flip state \( \psi
^{1,n}(\tilde{\alpha }) \)
is thus a ground state for \( N_{e}=N_{d}-n+1 \) electrons.

The proof presented here is considerably simpler than the proof in
\cite{Mielke93}.
Compared to the proof in \cite{Mielke99} it has the advantage that the
existence
of the single-spin flip ground state is trivial, whereas in \cite{Mielke99}
a lengthy calculation (hidden in fotenote 13) was necessary to show that. On
the other hand, for a translationally invariant multi-band system the basis
used here is clearly artifical. But if one uses a natural basis of Bloch
states,
it is very hard to construct the single spin-flip states from multi spin-flip
states.

\section{Summary and Outlook}

In this paper it is shown that for a general Hubbard model with degenerate
single
particle ground states local stability of ferromagnetism implies global
stability
of ferromagnetism. To be more precise: If there are no single spin-flip ground
states, all ground states have the maximal spin. This result holds if the
number
of electrons is less than or equal to the number of degenerate single particle
ground states. A similar result has been proven for a single band Hubbard model
in \cite{Mielke99} and the present result can be seen as a generalization.
It holds for a Hubbard model with more than one band, it holds even if the
model
does not have translational invariance. Furthermore, the proof presented here
is much simpler than the proof in \cite{Mielke99}.

The result is important since in many cases it is much simpler to show local
stability of ferromagnetism than global stability. In \cite{Tasaki96} showed
that under very general conditions the flat band ferromagnetism can be extended
to situations where the lowest band is not flat but has a weak dispersion. He
was able to prove that in such a situation the ferromagnetic ground state is
locally stable. It would be very interesting to obtain conditions under which
in that case the ferromagnetic ground state is globally stable, i.e. where it
is the real ground state of the system. This is clearly a very difficult
project.
The present theorem does not apply since Tasakis model does not have degenerate
single particle ground states. But one may hope that a generalization is
possible.
If in a situation with degenerate single particle ground states the
ferromagnetic
ground state is the only one, it is possible that ferromagnetism is stable with
respect to small perturbations of the Hamiltonian. The main problem is clearly
that for a general model there is no gap in the single particle spectrum like
in the models discussed by Tasaki \cite{Tasaki96}.

{}From a physics point of view ferromagnetism for models with a partially flat
band, as studied in this paper, differs from the flat-band ferromagnetism,
since
a flat-band ferromagnet is typically an insulator, whereas models with a
partially
flat band describe metals. Therefore our new approach is a step towards the
understanding of metallic ferromagnetism in the Hubbard model.

\subsection*{Acknowledgement}

I am grateful to Hideo Aoki for drawing my attention on refs. \cite{Arita}
and \cite{Arita98}.

\end{document}